\documentclass[aps,prl,showpacs,twocolumn,amsmath,amsfonts,amssymb,superscriptaddress]{revtex4-1}
\usepackage{graphicx}
\usepackage[tight]{subfigure}
\usepackage{color,soul}
\usepackage{appendix}

\usepackage{times}
\usepackage{float}
\usepackage{latexsym,amsmath,amssymb,bm,euscript}
\usepackage{color}
\usepackage{epstopdf}
\usepackage[colorlinks=true,linkcolor=blue,citecolor=blue]{hyperref}

\hypersetup{
    bookmarks=true,         % show bookmarks bar?
    unicode=false,          % non-Latin characters
    pdftoolbar=true,        % show Acrobat
    pdfmenubar=true,        % show Acrobat
    pdffitwindow=false,     % window fit to page when opened
    pdfstartview={FitH},    % fits the width of the page to the window
    pdftitle={My title},    % title
    pdfauthor={Author},     % author
    pdfsubject={Subject},   % subject of the document
    pdfcreator={Creator},   % creator of the document
    pdfproducer={Producer}, % producer of the document
    pdfkeywords={keyword1} {key2} {key3}, % list of keywords
    pdfnewwindow=true,      % links in new window
    colorlinks=true,       % false: boxed links; true: colored links
    linkcolor=magenta, %red,          % color of internal links (change box color with linkbordercolor)
    citecolor=blue,        % color of links to bibliography
    filecolor=magenta,      % color of file links
    urlcolor=cyan           % color of external links
}

\newcommand{\beq}{\begin{equation}}
\newcommand{\eeq}{\end{equation}}
\newcommand{\ba}{\begin{array}{ccc}}
\newcommand{\ea}{\end{array}}

\def\bea{\begin{eqnarray}}
\def\eea{\end{eqnarray}}
\newcommand{\bml}{\begin{multline}}

\newcommand{\eeqm}{\end{multline}}
\newcommand{\bsp}{\begin{split}}
\newcommand{\esp}{\end{split}}

\DeclareMathOperator{\Tr}{Tr}

\usepackage[normalem]{ulem} % to strike things out

\begin{document}

\title{Entanglement signatures of emergent Dirac fermions:\\
kagome spin liquid \& quantum criticality}

\author{Wei Zhu}
\email{weizhu@lanl.gov}
\affiliation{Theoretical Division, T-4 \& CNLS, Los Alamos National Laboratory, Los Alamos, New Mexico 87545, USA}

\author{Xiao Chen}
\email{xchen@kitp.ucsb.edu}
\affiliation{Kavli Institute for Theoretical Physics,
University of California at Santa Barbara, CA 93106, USA}

\author{Yin-Chen He}
\email{yinchenhe@perimeterinstitute.ca}
\affiliation{Department of Physics, Harvard University, Cambridge, Massachusetts, 02138, USA}

\author{William Witczak-Krempa}
\email{w.witczak-krempa@umontreal.ca}
\affiliation{D\'epartement de physique, %and Regroupement qu\'eb\'equois sur les mat\'eriaux de pointe,
Universit\'e de Montr\'eal, Montr\'eal (Qu\'ebec), H3C 3J7, Canada}

\begin{abstract}
Quantum spin liquids (QSL) are exotic phases of matter that host fractionalized excitations.
%\new{which originates from long-range entanglement}.
%\old{Local probes are insufficient to characterize them
%whereas quantum entanglement can serve as a diagnostic tool due to its non-locality}.
% Since the underlying physics is root in long-ranged quantum entanglement,
% local probes are hardly capable of characterizing them whereas quantum entanglement can serve as a diagnostic tool due to its non-locality.
It is difficult for local probes to characterize QSL, whereas quantum entanglement can serve as a powerful diagnostic tool due to its non-locality.
The kagome antiferromagnetic Heisenberg model is one of the most studied and experimentally relevant models for QSL, but its solution remains under debate.
Here, we perform a numerical Aharonov-Bohm experiment on this model and uncover universal features of the entanglement entropy. By means of the density-matrix renormalization group, we reveal
the entanglement signatures of emergent Dirac spinons, which are the fractionalized excitations of the QSL.
This scheme provides qualitative insights into the nature of kagome QSL, and can be used to study other quantum states of matter. As a concrete example, we also benchmark our methods on an interacting quantum critical point
between a Dirac semimetal and a charge ordered phase.
\end{abstract}

\date{\today}
\maketitle
Quantum spin liquids (QSLs)
are highly entangled states of matter with exotic excitations behaving as fractions of fundamental particles \cite{SavaryBalentsReview,Zhou2017}.
In the vigorous search for candidate materials, herberthsmithite ranks as one of the most promising ones \cite{Norman2016}.
Although it displays several signatures of a QSL, consensus between experimental
and theoretical studies is hindered not only by disorder, but also by the lack of understanding for the minimal model \cite{Norman2016}.
A starting point for a theoretical description of this correlated material is the antiferromagnetic Heisenberg model on the kagome lattice,
which is built out of corner sharing triangles that frustrate the anti-alignment of spins favored on each bond.
Frustration renders this model difficult to solve
~\cite{Sachdev1992,Hastings2000,Ran2007,Hermele2008,Evenbly2010,Iqbal2013,Yan2011,Depenbrock2012,Jiang2012,Mei2016,Jiang2016,Liao2017,YCHe2017},
and debates between different theoretical scenarios persist. For instance,
in a numerical study using the density-matrix renormalization group (DMRG)~\cite{White1992},
a gapped groundstate without magnetic order was found~\cite{Yan2011}.
More recent numerical studies \cite{Jiang2016,Liao2017,YCHe2017} suggest a gapless QSL. In this direction, evidence for a Dirac QSL~\cite{Hastings2000,Hermele2008} was obtained using DMRG simulations~\cite{YCHe2017}.

The fractional excitations of a QSL cannot be characterized by local order parameters.
Instead, entanglement may directly reveal fractionalization by virtue of its non-local nature.
For example, the so-called topological entanglement entropy can be used to detect gapped QSL (e.g.\ see the review~\cite{Eisert2010}).
Less comprehensive but nevertheless interesting results for the entanglement properties of gapless systems were obtained~\cite{Fradkin2006,Max2009,Stephan2009,YiZhang2011,Ju2012,Casini2009,XiaoChen2015}.
For example, attempts have been made to understand the entanglement response to a flux insertion~\cite{Max2009,XiaoChen2017,Whitsitt2017}.
However, our understanding of interacting systems remains limited, especially for realistic QSL models.

\begin{figure*}[t]
\includegraphics[width=0.85\textwidth]{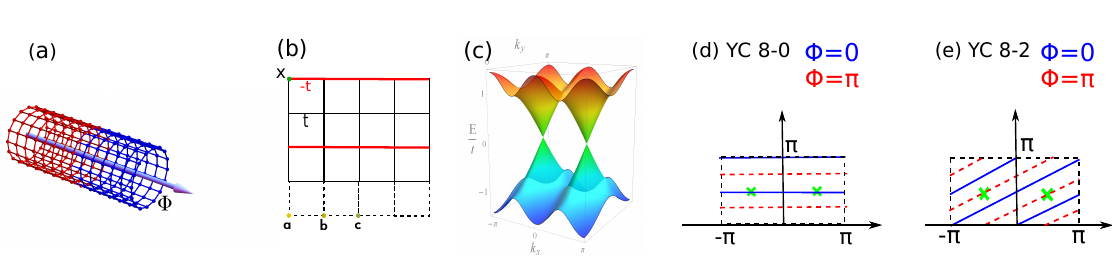}\\
\includegraphics[width=0.85\textwidth]{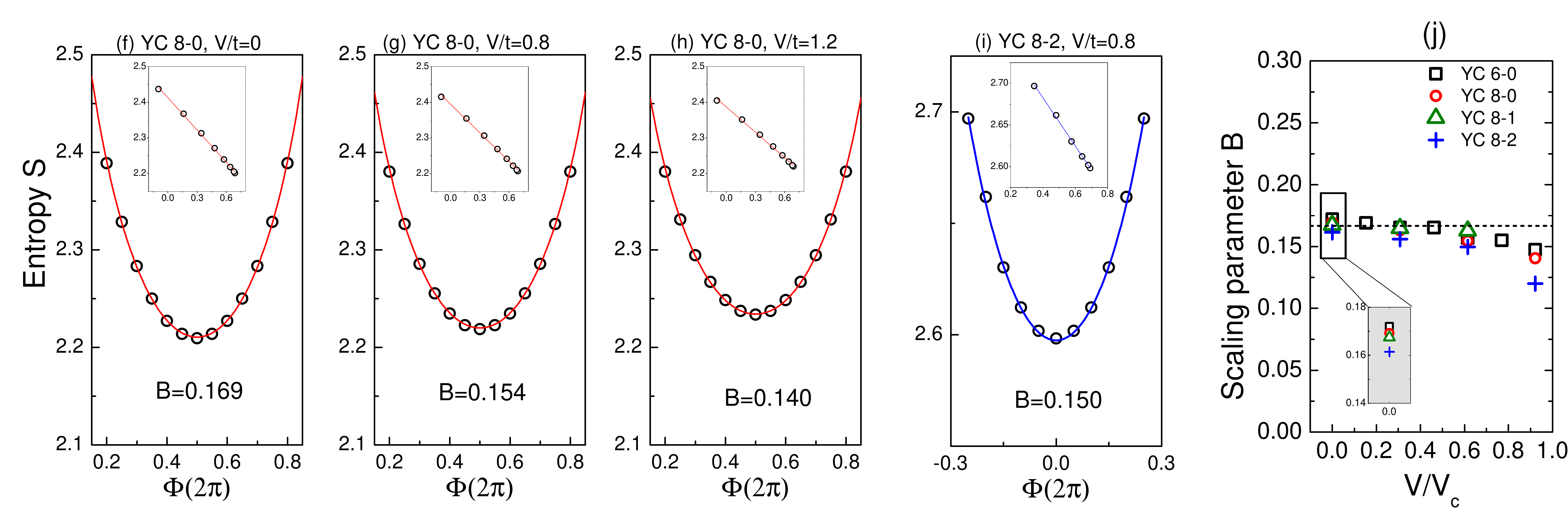}
\caption{{\bf Entanglement of quantum critical Dirac fermions.}
{\bf (a)} Cylinder with a flux insertion.
{\bf (b)} $\pi-$flux square lattice model.
The different cylinder types, YC$n$-0, YC$n$-1, YC$n$-2,
correspond to identifying $x$ with $a, b$, or $c$, respectively.
{\bf (c)} The energy dispersion at $V=0$ shows two Dirac cones at $(\pm\tfrac{\pi}{2},\tfrac{\pi}{2})$.
{\bf (d-e)} Blue (red) lines shows allowed momenta for $\Phi\!=\!0$ ($\pi$) in the Brillouin zone for an infinite cylinder with circumference $L_y\!=\! 8$.
The green crosses show the positions of the Dirac points.
{\bf (f)-(i)} EE versus flux $\Phi$ on a YC8-0 infinite cylinder
for $V/t=0, 0.8, 1.2$. {\bf (h)} EE for the YC8-2 cylinder at $V/t\! =\! 0.8$.
In (f-i), the lines are the best fits to Eq.(\ref{eq:scaling2});
insets show the data plotted in terms of the $\Phi$-dependent part of \eqref{eq:scaling2} and the linear fits.
{\bf (j)} Fitting parameter $B$ versus the repulsion strength $V/V_c$ for various cylinder types. The dashed line is
the prediction for non-interacting Dirac fermions.
} \label{fig:sq_scaling}
\end{figure*}

Here we investigate the quantum entanglement of the QSL in the kagome antiferromagnetic Heisenberg model in response to a magnetic flux.
We perform large-scale DMRG simulations (see Methods)
on infinitely long cylinders through which the flux is threaded, see Fig.~\ref{fig:sq_scaling}(a).
The main finding is that the entanglement entropy (EE) is highly sensitive to the flux
and is consistent with emergent Dirac cones for fractionalized spinon excitations.
This constitutes new evidence that the gapless Dirac QSL is the groundstate of the kagome antiferromagnetic Heisenberg model.
Moreover, in order to illustrate these entanglement signatures in a simpler setting,
we will begin by studying a strongly interacting quantum  phase transition
between a Dirac semimetal and a charge ordered state.
These results not only help with the interpretation of the data for the debated kagome QSL,
they also shed new light on quantum critical states of matter.

\textbf{Entanglement scaling of Dirac fermions.}
We consider a general quantum system on an infinitely long cylinder, and we calculate the von Neumann EE $S$ of the groundstate
by partitioning the system into two halves as shown in Fig.~\ref{fig:sq_scaling}(a).
$S$ quantifies the amount of quantum entanglement between the halves,
and takes the form \cite{Eisert2010}: $S = \alpha \frac{L_y}{a} - \gamma + \cdots$,
where $a\ll L_y$ is a microscopic scale such as a lattice spacing.
The first term arises for the groundstates of most Hamiltonians,
and is called the ``boundary law'' because it scales with the length of the partition, the circumference $L_y$ of the cylinder.
The boundary law term is of little interest in itself
because it is not universal, as it depends on the microscopic scale $a$. In contrast, the subleading term $\gamma$ is a low energy  property
 and does not depend on $a$.
Additional information can be extracted by inserting a flux $\Phi$ in the cylinder,
and studying the response of $\gamma$ to $\Phi$.
The flux dependence can then be used as fingerprint to identify the quantum state \cite{Max2009,XiaoChen2017}.

For a two-component free (gapless) Dirac fermion on the cylinder, the EE takes the following form in the continuum
\begin{equation}\label{eq:scaling1}
S =\alpha\frac{L_y}{a} - B \ln\big|2\sin\big(\frac{\Phi}{2}\big)\big|,
\end{equation}
where $B\!=\! 1/6$ \cite{XiaoChen2017,Whitsitt2017}.
One way to understand the above scaling behavior is to realize that the transverse momenta are quantized on a cylinder, and the flux $\Phi$ will move those quantized momenta towards or away from the Dirac point.
Intuitively, the subleading term quantifies how far the quantized momenta are from the Dirac point.
When $\Phi\to 0, 2\pi$, one momentum exactly hits the Dirac point leading to a diverging $\gamma$.
When $\Phi=\pi$, the momenta are farthest away from the Dirac point so $S$ becomes minimal.

In order to compare the Dirac scaling function \eqref{eq:scaling1} with a generic interacting system on a lattice, we
need to modify it as follows:
\begin{equation}\label{eq:scaling2}
S = \alpha \frac{L_y}{a} - B \sum_{n=1}^N  \ln\left| 2\sin\left[\frac{1}{2}(s\Phi - \Phi^c_n) \right] \right|\,.
\end{equation}
There are 3 new ingredients.
First, we can have $N>1$ different Dirac fermions.
Second, the momenta of the corresponding Dirac points in the Brillouin zone can be different, which is encoded in the shift $\Phi_n^c$.
$\Phi_n^c$ is proportional to the flux at which the $n{\textrm{th}}$ Dirac fermion's gap vanishes on the cylinder; $1/s$ is the proportionality constant.
Third, the Dirac fermions can carry a fractional charge $s$ (e.g.\ $s=\pm 1/2$ in the kagome QSL), hence they will feel a flux $s\Phi$ instead of $\Phi$.
The flux response of $S$ thus gives a clear way to identify fractionalization, which is notoriously difficult using conventional approaches.

\textbf{Quantum criticality.}
We start our discussion by explaining the salient entanglement features of a quantum critical transition between a Dirac semimetal
and an interaction-driven insulator with charge order.
By virtue of its universality, such a transition is relevant in contexts such
as charge density wave transitions in graphene \cite{Igor2006}.
We consider the $\pi-$flux square lattice model with a short-ranged repulsion $V$:
\begin{equation}\label{eq:sq}
  H=t\sum_{\langle ij\rangle} \left( (-1)^{s_{ij}} c^\dagger_i c_j +{\rm h.c.}\right) + V\sum_{\langle i j\rangle} n_i n_j,
\end{equation}
where $c^\dagger_i$ is the creation operator for a spinless fermion on site $i$, and $n_i$ is the particle number operator.
The phase factor $(-1)^{s_{ij}}$ generates a $\pi$ flux on each square plaquette as shown in Fig.~\ref{fig:sq_scaling}(b).
In the non-interacting limit, $V\!=\! 0$, the band structure hosts two Dirac cones located at $(k_x,k_y)=(\pm \pi/2,\pi/2)$,
as shown in Fig.~\ref{fig:sq_scaling}(c).
The repulsive interaction between nearest-neighbors drives a quantum phase transition from the Dirac semimetal to
a charge density wave phase through the strongly interacting Gross-Neveu-Yukawa
quantum critical point \cite{Igor2006},
where the Dirac quasiparticles are destroyed by quantum fluctuations.
Numerical studies studying local observables were performed~\cite{LeiWang2014,ZXLi2015},
but the entanglement properties near the critical point have not been investigated.

Fig.~\ref{fig:sq_scaling} shows the EE $S$ of model (\ref{eq:sq}) on infinite cylinders threaded by a flux $\Phi$.
First, we find that $S(\Phi)$ agrees with the scaling function \eqref{eq:scaling2}.
Specifically, we have two (non-fractionalized) Dirac fermions: $N\!=\!2$ and $s\!=\!1$.
For one type of cylinder (YC8-0, see the caption of Fig.~\ref{fig:sq_scaling}), the momenta are quantized such that they hit the Dirac points when $\Phi\!=\!0$. Therefore, $\Phi_{1,2}^c=0$ in the scaling function \eqref{eq:scaling2}.
For the other type of cylinder (YC8-2), the quantized momenta hit the Dirac points when $\Phi=\pi$ hence we have $\Phi_{1,2}^c=\pi$.

The scaling behavior is robust in the entire Dirac semimetal phase $V<V_c\approx 1.3t$ (Fig.~\ref{fig:sq_scaling}(f-i)),
despite the decrease of the prefactor $B$ as the quantum critical point is approached.
In the charge ordered phase $V\!>\!V_c$, the entropy does not follow the scaling behavior \eqref{eq:scaling2}
anymore (see \cite{SM}).
Finally, we emphasize that the scaling behavior is robust against changes to the circumference and cylinder type.
As shown in Fig.~\ref{fig:sq_scaling}(j), for various cylinder sizes and types,
the scaling parameter $B$ follows the same decreasing trend as the critical point is approached.
As a consistency check, we observe that $B$ approaches its non-interacting value $B\!=\! 1/6$
when $V\!\to\!0$.
In order to confirm that the above properties of the EE are universal, we analyzed another model on the honeycomb lattice and reached identical conclusions~\cite{SM}.

The deviation of $B$ from the expectation of free theory is a consequence of the increasing quantum fluctuations as the critical point is approached.
To explain this fact we now invoke a field theory description.
In order to make this theory tractable,
we extend the number of Dirac fermions to $N\!\gg \!1$.
In this limit, it was recently shown \cite{Whitsitt2017} that the subleading correction $\gamma$ vanishes at leading order.
This represents a drastic
reduction compared with a weakly interacting Dirac semimetal, where $\gamma$ is directly proportional to the number of
Dirac fermions, $N$.
Extrapolating to finite $N$, we conjecture that as the quantum critical point of Eq.~\eqref{eq:sq} is approached, $B$ is suppressed.
Our data in Fig.~\ref{fig:sq_scaling} corroborates this conclusion.

\begin{figure}[b]
\includegraphics[width=0.49\textwidth]{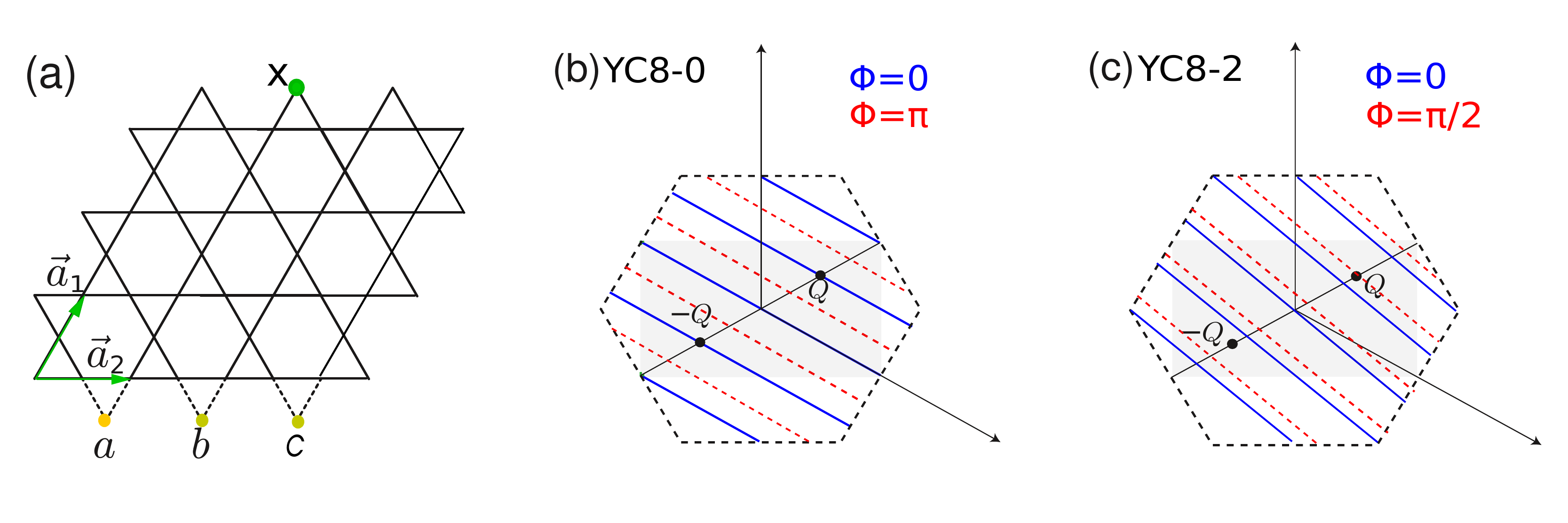}
\caption{
{\bf Kagome cylinders \& allowed momenta.} {\bf (a)} The different types of kagome cylinders, YC8-0, YC8-2 or YC8-4,
correspond to identifying site $x$ with site $a$, $b$, or $c$, respectively.
The blue (solid) and red (dashed) lines show the allowed momenta
for an infinite cylinder of type {\bf (b)} YC8-0 and {\bf (c)} YC8-2. The gray rectangle is the
magnetic Brillouin zone due to the $\pi$-fluxes in the hexagons.
The two Dirac points of the QSL are at $\pm Q=\pm(\pi/2,\pi/2)$.
 } \label{fig:kag_BZ}
\end{figure}

\begin{figure} %[!htb]
\includegraphics[width=0.51\textwidth]{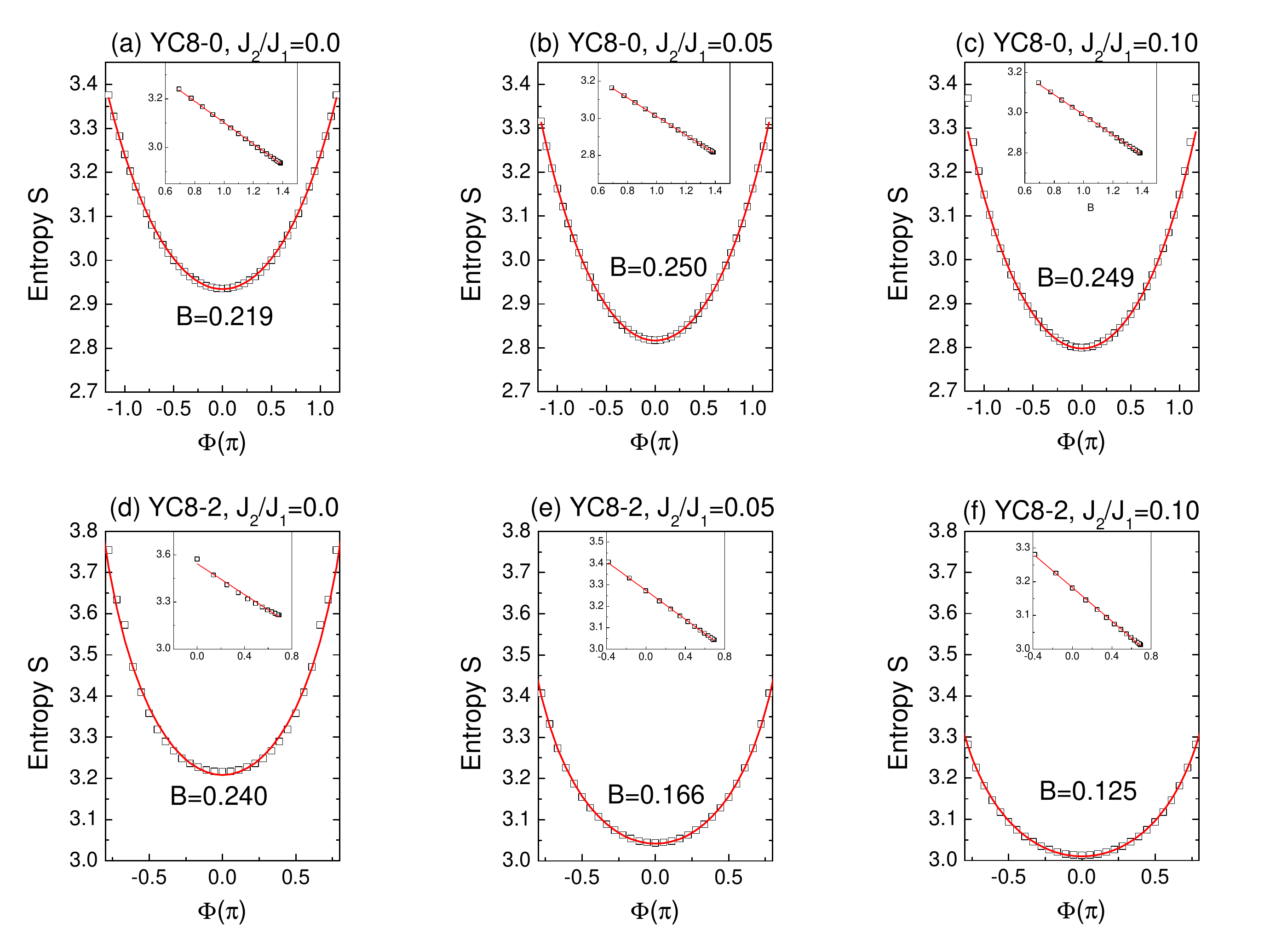}
\caption{{\bf Entanglement entropy for the kagome Heisenberg model.} DMRG results for the EE versus the flux $\Phi$ for the spin-$1/2$ kagome antiferromagnetic Heisenberg model on an infinite cylinder with $8$ sites in the periodic direction.
{\bf (a-c)} are for the YC8-0 cylinder, and {\bf (d-f)} are for the YC8-2 cylinder.
The red lines are best fits to Eq.~\eqref{eq:scaling2}.
{\bf Insets}: The EE plotted as function of the $\Phi$-dependent part of \eqref{eq:scaling2} and the corresponding linear fits.
} \label{fig:kag_scaling}
\end{figure}

\textbf{Kagome spin liquid.}
We now tackle our main objective, the spin-1/2 antiferromagnetic Heisenberg model on the kagome lattice (Fig.~\ref{fig:kag_BZ}):
\begin{equation}\label{eq:kag}
  H= J_1\sum_{\langle ij\rangle } \mathbf S_i \cdot \mathbf S_j + J_2\sum_{\langle\!\langle ij \rangle\!\rangle } \mathbf S_i \cdot \mathbf S_j,
\end{equation}
where $J_1,J_2>0$ are nearest and next-to-nearest neighbor antiferromagnetic couplings, respectively.
Although we focus on the $J_2\!=\!0$ case, we shall also consider the effects of a small $J_2/J_1$, which
makes the numerical results more stable.
Figure~\ref{fig:kag_scaling} shows the flux dependence of the EE at $J_2/J_1=0,0.05, 0.1$,
as well as for two cylinders types, YC8-0 and YC8-2.
We first note that in all cases $S$  strongly depends on $\Phi$,
which is a hallmark of low energy excitations. In contrast, a state with a large gap would be essentially insensitive to $\Phi$.
Importantly, the data in Fig.~\ref{fig:kag_scaling} can be accurately fitted with the scaling function \eqref{eq:scaling2}.
The parameters ($N$, $s$, $\Phi_n^c$) are chosen to match the $\pi$-flux Dirac QSL~\cite{Hermele2008,Hastings2000},
in which 4 two-component Dirac spinons ($N\!=\!4$ accounts for the spin and valley degrees of freedom) carry fractionalized spin $s\!=\!1/2$.
The shifts $\Phi_n^c$ depend on the cylinder type (Fig. \ref{fig:kag_BZ}), and are given in Table~\ref{table:cone}.

In Fig.~\ref{fig:kag_scaling}, we observe that the scaling function \eqref{eq:scaling2} accurately fits the data for all the geometries and couplings considered.
When $J_2\!=\!0$, the fitting parameter $B$ takes the value $0.219$ on YC8-0 and $0.240$ on YC8-2.
This is larger than the free fermion value $B=1/6$.
This deviation from the non-interacting value can be understood by the fact that the low energy description of the Dirac QSL is
in terms of Dirac spinons strongly coupled to an emergent photon \cite{Hermele2008}.
In contrast, in the large-$N$ limit, when the gauge fluctuations become suppressed,
the leading order EE is that of $N$ free Dirac fermions~\cite{SM}. Our data suggests that the value of $B$ becomes renormalized at finite $N$ but the flux dependence remains largely unchanged.

\begin{table}[b]
\caption{\label{table:cone} Values of the shifts $\Phi_n^c$ in Eq.~\eqref{eq:scaling2} for the kagome model.
On the YC8-0 cylinder, $\Phi_n^c$ equals the internal gauge flux $\phi\!=\!\pi$. }
\setlength{\tabcolsep}{0.25cm}
\renewcommand{\arraystretch}{1.4}
\begin{tabular}{cccc}
\hline\hline
Dirac flavor &      & YC8-0 & YC8-2   \\ \hline
$\uparrow, Q$ &$\Phi_1^c$ & $\pi$ &  $\pi/2$  \\ \hline
$\uparrow, -Q$ &$\Phi_2^c$ & $\pi$ &  $-\pi/2$ \\ \hline
$\downarrow, Q$ &$\Phi_3^c$ & $\pi$ &  $-\pi/2$ \\ \hline
$\downarrow, -Q$ &$\Phi_4^c$ & $\pi$ &  $\pi/2$ \\
\hline \hline
\end{tabular}
\end{table}

At $J_2/J_1=0.05$ and $0.1$, we see that the fits are more accurate. $B$ increases slightly to $0.250$ for YC8-0.
On YC8-2, $B$ first decreases to $0.166$ then to $0.125$. This tendency for $B$ to be smaller on type 2 cylinders was
seen above for the interacting Dirac semimetal on the square lattice. Here, the finite size effects for YC8-2 may be
further amplified compared to YC8-0.
One reason may be that the allowed momentum lines are always closer to the gapless Dirac points on the YC8-2 cylinder.
As $J_2/J_1$ is increased, a conventional ordered state becomes favored and the resulting phase transition is expected to
leave an imprint on the entanglement. A more detailed analysis in this direction will likely lead to new insights
into the properties of the kagome QSL and its phase transitions.

\textbf{Conclusions.}
By monitoring the entanglement entropy response to a flux threaded in a cylindrical geometry, we were
able to gain new insights about two physical systems: 1) a quantum critical phase transition of itinerant electrons and 2) the frustrated kagome Heisenberg model.
In the first case, the entanglement entropy tracks the evolution of the Dirac fermions as the quantum critical point is approached.
For the kagome model, the flux dependence of the entanglement entropy
unambiguously points to four emergent Dirac cones of fractionalized excitations (spinons). The robust
features we have identified on
various cylinder types and values of the Heisenberg couplings strongly suggest that the kagome Heisenberg model is a gapless Dirac QSL.
These new insights will help with the modeling of candidate materials such as herbertsmithite.
Our two concrete examples give us confidence that entanglement signatures will become a valuable tool
in the investigation of a broad class of quantum states of matter.

\section{Methods} \label{methods}
The groundstates of models \eqref{eq:sq} and \eqref{eq:kag} were determined using the density-matrix renormalization
group (DMRG)~\cite{White1992}, which is a powerful algorithm to determine in an unbiased fashion the low-lying states of quantum systems. In our simulations, we work on infinitely long cylinders with a finite circumference \cite{McCulloch2008}.
We can reach circumferences of 4 unit cells on the kagome lattice, which is close to the current computational limit.
In our simulations, matrix product states of bond dimension
$1600$ were sufficient to describe the entanglement entropy of the $\pi$-flux model on the square lattice, whereas
for the kagome model, a bond dimension of $6000$ was used, as the subsystem
entanglement is significantly larger.
The numerical flux insertion experiment was performed by adiabatically
changing (twisting) boundary conditions in the Hamiltonian.
In the simulations, we impose twisted boundary conditions along the circumference of the cylinder
by replacing the terms $c^\dagger_i c_j +\mbox{h.c.}$ ($S^+_i S^-_j+\mbox{h.c.}$) for all bonds
crossing the $y$-boundary with
$c^\dagger_i c_j e^{i\Phi}+\mbox{h.c.}$ ($S^+_i S^-_j e^{i\Phi}+\mbox{h.c.}$) \cite{He2014a}.

Once the groundstate $|\Psi(\Phi)\rangle$  is computed, we partition the cylinder into two halves, $A$ and $B$,
and calculate the von Neumann entanglement entropy $S(\Phi)=-\sum_i \lambda_i(\Phi) \ln\lambda_i(\Phi)$,
where $\lambda_i$ are the eigenvalues of reduced density matrix of the $A$ half, $\rho_A(\Phi)=\Tr_B |\Psi(\Phi)\rangle \langle \Psi(\Phi)|$.
The entanglement entropy measures the amount of quantum entanglement between a region $A$ and its complement.
In order to obtain the entanglement entropy at different $\Phi$, we used an adiabatic scheme:
the groundstate $|\Psi(\Phi)\rangle$ is taken as the initial state for the calculation at $\Phi+\Delta\Phi$.

The fit of $S(\Phi)$ is based on the least-squares method. The data points near the entropy minimum are used in the fitting
as these are the most reliable.
For the kagome lattice, the entropy in the range $|\Phi|<0.24\pi$ was used for the fitting process.
For the square lattice, the entropy in the range $|\Phi-\Phi^{\rm min}|<0.4\pi$ was used in the fits,
where $\Phi^{\rm min}$ is flux value where the entropy is minimal.
We have verified that all of the fits are stable and independent of the data range we select,
except for the YC8-2 $J_2=0$ case (Fig.~\ref{fig:kag_scaling}(a)).
For YC8-2 $J_2=0$, the scaling parameter $B$ could vary from $0.20$ to $0.25$, as we change the data regime from
$|\Phi|<0.2\pi$ to $|\Phi|<0.4\pi$.
This can be attributed to the stronger finite-size effects at this coupling on the YC8-2 cylinder.

\textbf{Acknowledgements.}
We are grateful for discussions with Hitesh J.~Changlani, Eduardo Fradkin, Joseph Maciejko, Subir Sachdev, Chong Wang and Seth Whitsitt. YCH thanks M. Zaletel, M. Oshikawa and F. Pollmann for previous collaboration on a related project.
WZ was supported by the DOE National Nuclear Security Administration through the Los Alamos National Laboratory LDRD Program.
XC was supported by a postdoctoral fellowship from the Gordon and Betty Moore Foundation, under the EPiQS initiative, Grant GBMF4304, at the Kavli Institute for Theoretical Physics.
YCH is supported by the Gordon and Betty Moore Foundation under the EPiQS initiative, GBMF4306, at Harvard University.
WWK was funded by a Discovery Grant from NSERC, and by a Canada Research Chair.
The work was initiated at a Moore funding postdoc symposium in Aspen.
Part of the work was performed at the Aspen Center for Physics, which is supported by National Science Foundation grant PHY-1066293.

{\bf Author contributions}
X.C.\ and Y.-C.H.\ initiated the project, W.Z.\ and Y.-C.H.\ performed the DMRG simulations.
All authors contributed equally to the analysis of the data and writing of the manuscript.

\bibliographystyle{apsrev4-1}
\bibliography{gapless,spin_liquid}

\clearpage
\begin{widetext}

\appendix
\begin{appendices}

%\tableofcontents

%\clearpage
\section{Entanglement entropy for free Dirac fermions on the cylinder}
The continuum Hamiltonian of Dirac fermions on the infinite cylinder reads
\begin{equation}
H=\sum_{k_y}\int \frac{dk_x}{2\pi}\Psi^{\dag}(\vec k)\begin{pmatrix}
m & k_x-i k_y \\
k_x+i k_y & -m
\end{pmatrix}\Psi(\vec k)
\label{H_Dirac}
\end{equation}
where $\Psi(k)=(\psi_1(k),\psi_2(k))^T$ is a two-component spinor and $m$ is the fermion mass. Here, $y$ is compact with periodicity $L_y$.
The transverse momentum $k_y$ takes discrete values, $k_y=\frac{2\pi n_y+\Phi}{L_y}$, where $\Phi$
is the flux inserted in the cylinder.

For an infinite cylinder bipartitioned into 2 semi-infinite cylinders (Fig.~\ref{fig:cylinder}), the EE for region $A$
(the left semi-infinite cylinder) obeys an area law
with a subleading term, $S=\alpha\, (L_y/\epsilon)-\gamma$. The subleading term, $-\gamma$, will be a function of the flux $\Phi$ inserted inside the cylinder \cite{Max2009,Arias2015,XiaoChen2017}.

We briefly review the computation of $\gamma$ by using the 1d decomposition method discussed in Ref.~\onlinecite{XiaoChen2017}. The Hamiltonian in Eq.~\eqref{H_Dirac} can be written as $H=\sum_{k_y} H^{\rm 1d}(k_y)$, where $H^{\rm 1d}(k_y)$ is a $(1+1)$-dimensional massive Dirac fermion.
For a semi-infinite interval, each  $H^{1d}(k_y)$ with an effective mass $\sqrt{m^2+k_y^2}$, contributes an EE~\cite{Cardy2004}
\begin{equation}
S^{\rm 1d}(k_y)=-\frac{1}{12} \ln\left[ (m^2+k_y^2)\epsilon^2 \right]
\end{equation}
%where the length of the interval is taken to be much larger than the inverse mass $L_A\geq 1/\sqrt{m^2+k_y^2}$;
$\epsilon$ is the short distance UV cutoff. %and $n$ is the R\'enyi index.
The total EE is then
\begin{equation}
S=-\frac{1}{12}\sum_{k_y}\ln\left[ (m^2+k_y^2)\epsilon^2 \right]
\label{S_renyi}
\end{equation}
%To evaluate the above expression, we need to regularize this infinite sum by using the generalized Zeta function regularization method,
%\begin{align}
%g(\lambda, mL_y)&=\sum_{-\infty}^{\infty}\ln\left[ (mL_y)^2+(2\pi)^2(n_y+\lambda)^2  \right]\nonumber\\
%&=\ln\left[ 2\cosh(mL_y)-2\cos(2\pi\lambda)  \right]
%\label{g_sub}
%\end{align}
For the massless case $m=0$ we are interested in here, by using the Zeta function regularization method, we have
\begin{equation}
S=\alpha \frac{L_y}{\epsilon}-\gamma
\end{equation}
where the subleading term is equal to
\begin{equation} \label{eq:gam-free}
\gamma =\frac{1}{6} \ln\big|2\sin\Big(\frac{\Phi}{2}\Big)\big|
\end{equation}
\begin{figure}%[hbt]
\centering
%\vspace*{-0cm}
\includegraphics[scale=.45]{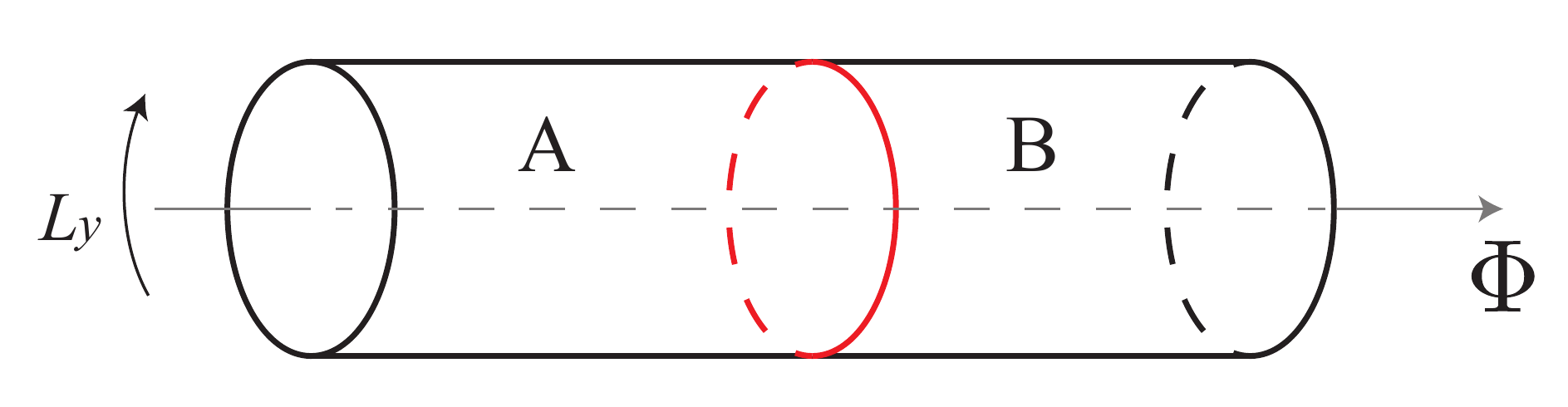}
\caption{Bipartition of an infinite cylinder, which is threaded by a flux $\Phi$.}
\label{fig:cylinder}
\end{figure}

When $m\neq 0$, the infinite sum in Eq.~\eqref{S_renyi} can also be obtained by using the generalized Zeta function regularization method,
\begin{eqnarray}
%g(\lambda, mL_y)
\sum_{k_y}\ln\left[ (m^2+k_y^2)L_y^2 \right]
&=\sum_{n_y = -\infty}^{\infty}\ln\left[ (mL_y)^2+(2\pi n_y+\Phi)^2  \right]\nonumber\\
&=\ln\left[ 2\cosh(mL_y)-2\cos\Phi \right]
\label{g_sub}
\end{eqnarray}
This result will be useful when we discuss the EE for the Gross-Neveu model in the next section.

\section{Entanglement entropy of interacting Dirac fermions at large $N$}
\subsection{Gross-Neveu quantum critical point}\label{ap:qft-gn}
Here we first briefly review the Gross-Neveu model in the large $N$ limit \cite{Gross1974}. The Euclidean Lagrangian for the Gross-Neveu model is
\begin{equation}
\mathcal L=-\bar{\Psi}_\alpha \partial \Psi_\alpha-\frac{g^2}{2N}(\bar{\Psi}_\alpha\Psi_\alpha)^2
\end{equation}
where the repeated flavor index $\alpha$ is summed over from $1$ to $N$, and $\partial=\Gamma_\mu \partial_\mu$ with
the 2-by-2 Gamma matrices $\Gamma_\mu$. The quartic interaction term $(\bar{\Psi}_\alpha\Psi_\alpha)^2$ can be decoupled by introducing a Hubbard-Stratonovich field $\phi$ and yields the Gross-Neveu-Yukawa Lagrangian:
\begin{align}
\mathcal L =-\bar{\Psi}_\alpha (\partial+\phi)\Psi_\alpha+\frac{N}{2g^2}\phi^2
\end{align}
After integrating out the fermions, the partition function $Z\!=\! \int D[\Psi]D[\phi]e^{-S}$ takes the form
\begin{align}
Z=\int D[\phi]\exp\left[N\Tr \ln (\partial+\phi)-\frac{N}{2g^2}\int d^3x \phi^2 \right]
\end{align}
In the large $N$ limit, the partition function can be evaluated using the saddle point method,
\begin{align}
\ln Z=N \Tr \ln (\partial+\phi)-\frac{N}{2g^2}\int d^3x\, \phi^2
\end{align}
Crucially, the saddle point configuration of the $\phi$ field is determined by solving the gap equation
\begin{align}
\frac{\langle\phi\rangle}{g^2}=\Tr G^F\!(x,x;\langle\phi\rangle)
\label{green_f}
\end{align}
where $G^F\!(x,x;\langle\phi\rangle)$ is fermionic Green's function. Thus, the fermions acquire a mass given by the saddle point value $\langle\phi\rangle$.
At the critical point, this mass vanishes on the infinite plane, but not on the cylinder. The mass will play a crucial role in the computation of the EE, as we shall see below.
In momentum space, the gap equation simplifies to
\begin{align}
\frac{\langle\phi\rangle}{g^2}=(\Tr \mathbb I)  \int \frac{d^3p}{(2\pi)^3}\frac{\langle\phi\rangle}{p^2+\langle\phi\rangle^2}
\end{align}
At the quantum critical point $\langle\phi\rangle$ vanishes, and the (non-universal) critical coupling is given by
\begin{align}
\frac{1}{g_c^2}= (\Tr \mathbb I) \int \frac{d^3p}{(2\pi)^3}\frac{1}{p^2}
\label{g_c}
\end{align}
where a momentum cutoff should be used.

%\subsubsection{2. Two-cylinder EE}
In order to compute the EE at the quantum critical point, we follow the calculation described in Ref.~\onlinecite{Whitsitt2017}, which makes use of the replica trick~\cite{Holzhey1994,Cardy2004}. % to compute EE for Gross-Neveu model at critical point.
The replica trick allows the calculation of the R\'enyi entanglement entropies, $S_n \!=\!\tfrac{1}{1-n}\ln\Tr\rho_A^n$, for integer values of the R\'enyi index $n$.
The analytic continuation of $S_n$ to $n\!=\! 1$, when possible, gives the von Neumann EE: $S=S_{n\to 1}$. The key identity is
\begin{align}
\Tr\rho_A^n=\frac{Z_n}{Z_1^n}
\end{align}
where $\rho_A$ is the reduced density matrix of region $A$, and $Z_n$ is the partition function defined over a special spacetime: an $n$-sheeted Riemann surface. The $n$ sheets are glued together at the boundary of region $A$, which in our case is a (flat) circle
dividing the cylinder in equal halves. We can formally evaluate the $n$-sheeted partition function for the Gross-Neveu-Yukawa model:
\begin{align}
\ln Z_n=N \Tr \ln (\partial_n+\langle\phi\rangle_n)-\frac{N}{2g_c^2}\int d^3 x\, \langle\phi\rangle_n^2
\end{align}
Around $n\approx 1$, we can expand the saddle point value of $\phi$, $\langle\phi(x) \rangle_n$, as
\begin{align}
\langle\phi(x) \rangle_n\approx m_1+(n-1)f(x)
\end{align}
where $m_1$ is the self-consistent mass satisfying the gap equation on the physical spacetime, $\Tr G^F_1\!(x,x;m_1)=m_1/g_c^2$,
and $f(x)$ is an unknown function on the Riemann surface. In the notation used above, $m_1=m$.
Therefore, $\ln Z_n$ can be written as
\begin{eqnarray}
  \ln Z_n=&N \Tr \ln(\partial_n+m_1)-\frac{N}{2g_c^2}\int d^3x_n m_1^2\nonumber\\
           &+(n-1)N \Tr \left(  \frac{f}{\partial_1 +m_1} \right)-(n-1)\frac{N}{g_c^2}\int d^3x\, m_1 f(x)
\end{eqnarray}
In the above expression, if we use Eq.~\eqref{green_f}, the last two terms will cancel each other. Therefore, we have
\begin{equation}
-\ln\frac{Z_n}{Z_1^n}=-N\left[ \Tr\ln(\partial_n+m_1)-n \Tr \ln(\partial_1+m_1) \right]
\end{equation}
This is the same result as for a free Dirac fermion with mass $m_1$. The mass $m_1$ can be obtained by solving the gap equation
at the critical point Eq.~\eqref{g_c}:
\begin{equation}
\frac{1}{L_y}\sum_{k_y}\int \frac{d^2p}{(2\pi)^2}\frac{1}{p^2+k_y^2+m_1^2}=-\frac{1}{4\pi L_y}\ln \left[ 2\cosh (m_1L_y)-2\cos\Phi \right]=0
\end{equation}
Notice that the above result is obtained by using Zeta function regularization which ignores the UV divergent term.
To satisfy the above equation, the mass becomes
\begin{equation}
m_1=\frac{1}{L_y}\mbox{arccosh}\left( \frac{1}{2}+\cos\Phi \right)
\end{equation}
If we plug the above expression into Eq.~\eqref{g_sub} (for the free Dirac fermion EE), we find that $\gamma=0$ for all values of $\Phi$.
Therefore, the subleading term is absent at leading order in $N$ for the Gross-Neveu model in the large $N$ limit. We expect $\gamma$ to become non-zero at next order in $N$, $\mathcal O(N^0)$.

\subsection{Quantum Electrodynamics (QED3)}
The Dirac QSL on the kagome lattice is described by a theory of Quantum Electrodynamics in 3 spacetime dimensions (QED3) in which 4 gapless Dirac fermions
are strongly coupled to an emergent gauge field, $a_\mu$. After extending the number of Dirac fermions to $N$, the Euclidean time Lagrangian
 becomes:
\begin{equation}
  \mathcal L = -\bar\Psi_\alpha (\partial+i a) \Psi_\alpha + \frac{1}{4e^2}f_{\mu\nu} f_{\mu\nu}
\end{equation}
where $ a=a_\mu \Gamma_\mu$ and $f_{\mu\nu}=\partial_\mu a_\nu-\partial_\nu a_\mu$ is the field strength tensor of the gauge field. The repeated flavor index $\alpha$ is again summed from 1 to $N$. Just as for the Gross-Neveu model, this theory is strongly interacting in the long-wavelength limit but becomes tractable at large $N$.\footnote{At large $N$, the theory
is in fact conformally invariant at low energy. Also, we can neglect the monopole operators that result from the
compactness of the gauge field in the lattice Hamiltonian.}
The leading order large-$N$ solution has the gauge field pinned to its saddle point value. However, in contrast
to $\langle\phi\rangle$ in the Gross-Neveu-Yukawa theory on the cylinder,
the saddle point value of $a_\mu$ vanishes. If present, such an expectation value would generate either a finite fermion density or current,
which does not happen on the cylinder (or infinite plane). The $n$-sheeted partition function is thus simply given by
\begin{equation}
  \ln Z_n = N \Tr\ln( \partial_n) + \mathcal O(N^0)
\end{equation}
This is the same answer as for $N$ free gapless Dirac fermions. One subtelty is that the internal gauge field can change the boundary conditions of the fermions
in order to lower the system's energy. This means that $\gamma$ in Eq.~\eqref{eq:gam-free} is replaced by
\begin{equation}
  \gamma = \frac{1}{6} \sum_\alpha \ln\big| 2\sin\Big(\frac{\Phi_\alpha^{\rm net}}{2}\Big)\big|
\end{equation}
where the net flux $\Phi_\alpha^{\rm net}$ felt by fermion $\alpha$ depends on both the external and internal fluxes.
This is discussed in more detail in the main text. % (for $N\!=\! 4$).
 At next order in $N$, the gauge fluctuations will contribute to $\gamma$.
Such a calculation is beyond the scope of the current paper, but it would be interesting in light of our DMRG results. For example, one would like
to know if the $1/N$ correction has the right sign to explain why the observed value exceeds the free Dirac fermion result, $B>1/6$.

\begin{figure}[t]
\includegraphics[width=0.45\textwidth]{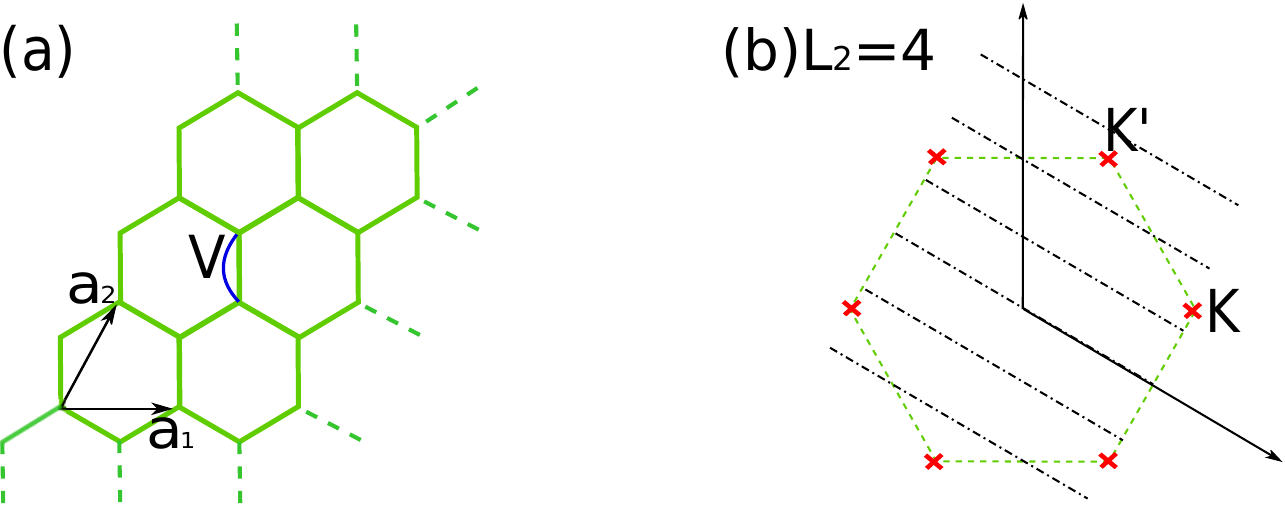}\\
\includegraphics[width=0.6\textwidth]{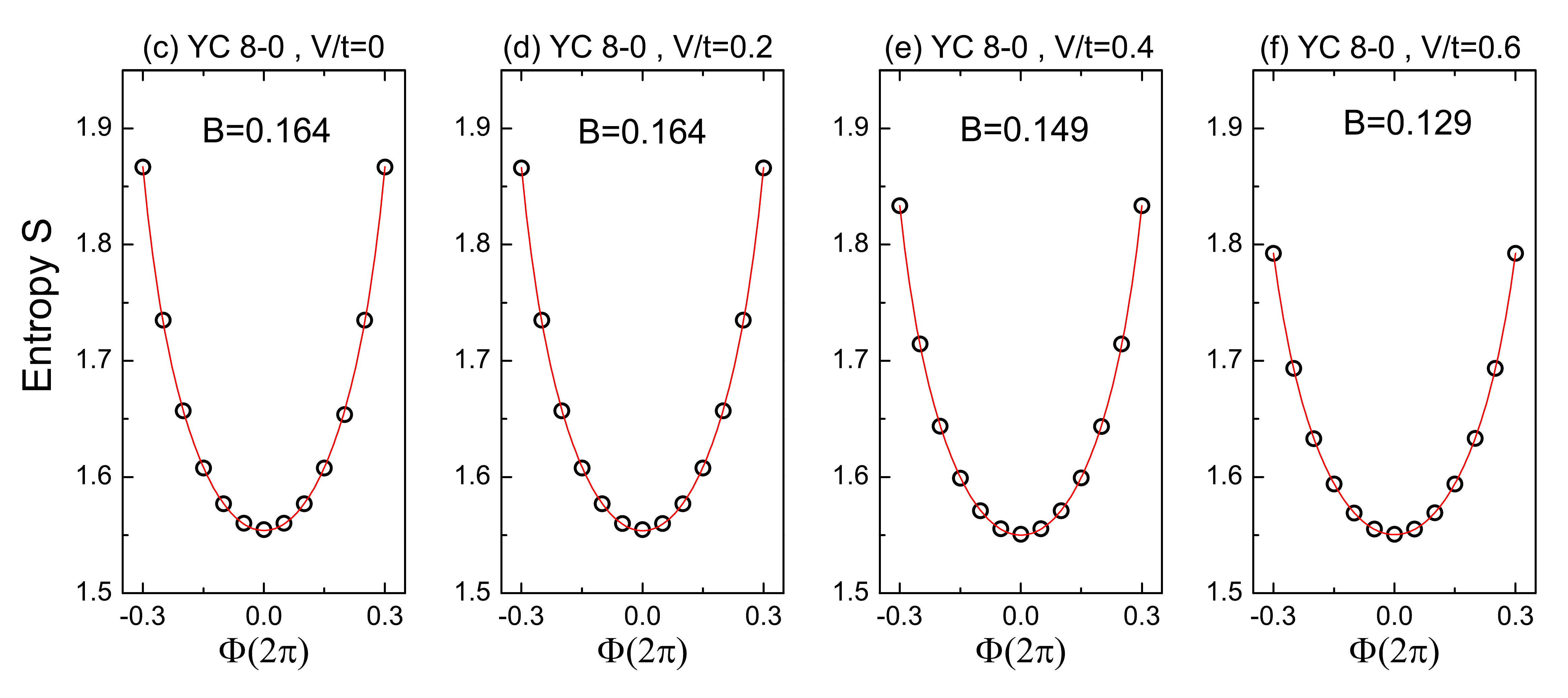}
\caption{{\bf Entanglement of quantum critical Dirac fermions on the honeycomb lattice.}
{\bf (a)} Honeycomb lattice on a cylinder with the compact direction being along $\vec a_2$.
{\bf (b)} Allowed momentum points (dot-dashed lines) in the Brillouin zone for an infinitely long cylinder with
circumference $L_2\!=\! 4$ (in unit of lattice vector $\vec a_2$). The red crosses mark the positions of the Dirac points.
{\bf (c-f)} The entanglement entropy versus the twist parameter $\Phi$ on an infinite cylinder with a
 circumference of $4$ unit cells, for various interactions
$V/t$. The red lines are best fits to the scaling function shown in the main text (Eq.~2).} \label{fig:hc}
\end{figure}

%\clearpage
\section{Quantum critical point of Dirac fermions on the honeycomb lattice}
In the main text, we have studied the fermionic quantum critical point of fermions in the $\pi-$flux square lattice model.
In order to confirm that the EE scaling observed is indeed universal, we analyze a different lattice model
that is expected to host a quantum critical point in the Gross-Neveu-Yukawa universality class.
The model is similar to the $\pi$-flux Hamiltonian (Eq.~3 of the main text) but defined instead on the honeycomb lattice.
The Hamiltonian contains hopping and repulsion terms:
\begin{equation}
  H = t\sum_{\langle ij\rangle}(c_i^\dag c_j+{\rm h.c.}) + V\sum_{\langle ij\rangle} n_i n_j
\end{equation}
% Another well-known example hosting Dirac cone structure is the honeycomb lattice, which
% can be realized in graphene thus is of great interest.
% To further confirm the universality of the entanglement entropy scaling behavior, in this section
% we simulated response of entropy to the external flux on the honeycomb lattice.
As in the main text, we perform large-scale DMRG simulations on infinite cylinders.
The transition from the Dirac semimetal at small $V/t$ to a charge density wave transition occurs at $V_c\simeq 1.36t$.
%should belong to the same universality class in the $\pi-$flux square lattice model.

As shown in Fig. \ref{fig:hc}(a-b), the allowed momenta of the $L_2=4$ (number of unit cells around the circumference)
cylinder do not hit the Dirac points ($\vec K$ and $\vec K'$) at zero flux,
thus the entanglement entropy has a minimum at $\Phi\!=\! 0$.
While the flux deviates from zero, entropy gradually increases, and becomes maximal approaching $\Phi_c=\pm \frac{2\pi}{3}$,
where the momentum lines hit a Dirac point.
In Fig.~\ref{fig:hc}(c-f), we fit the EE using the same scaling ansatz as in the main text.
The entanglement entropy dependence on the twisted boundary condition
perfectly matches the scaling function, for the whole Dirac semimetal phase.
At small $V/t$, the fitting prefactor $B$ is close to the value for free Dirac fermions, $B=1/6$, as expected.
As the repulsion is increased, $B$ decreases. This behavior was also observed in the square lattice model in the main text, and justified using
field theory (see Section~\ref{ap:qft-gn} and the main text). The agreement between the honeycomb and square lattice DMRG results
strongly suggest that our results probe universal low energy properties.

\begin{figure}[t]
\includegraphics[width=0.4\textwidth]{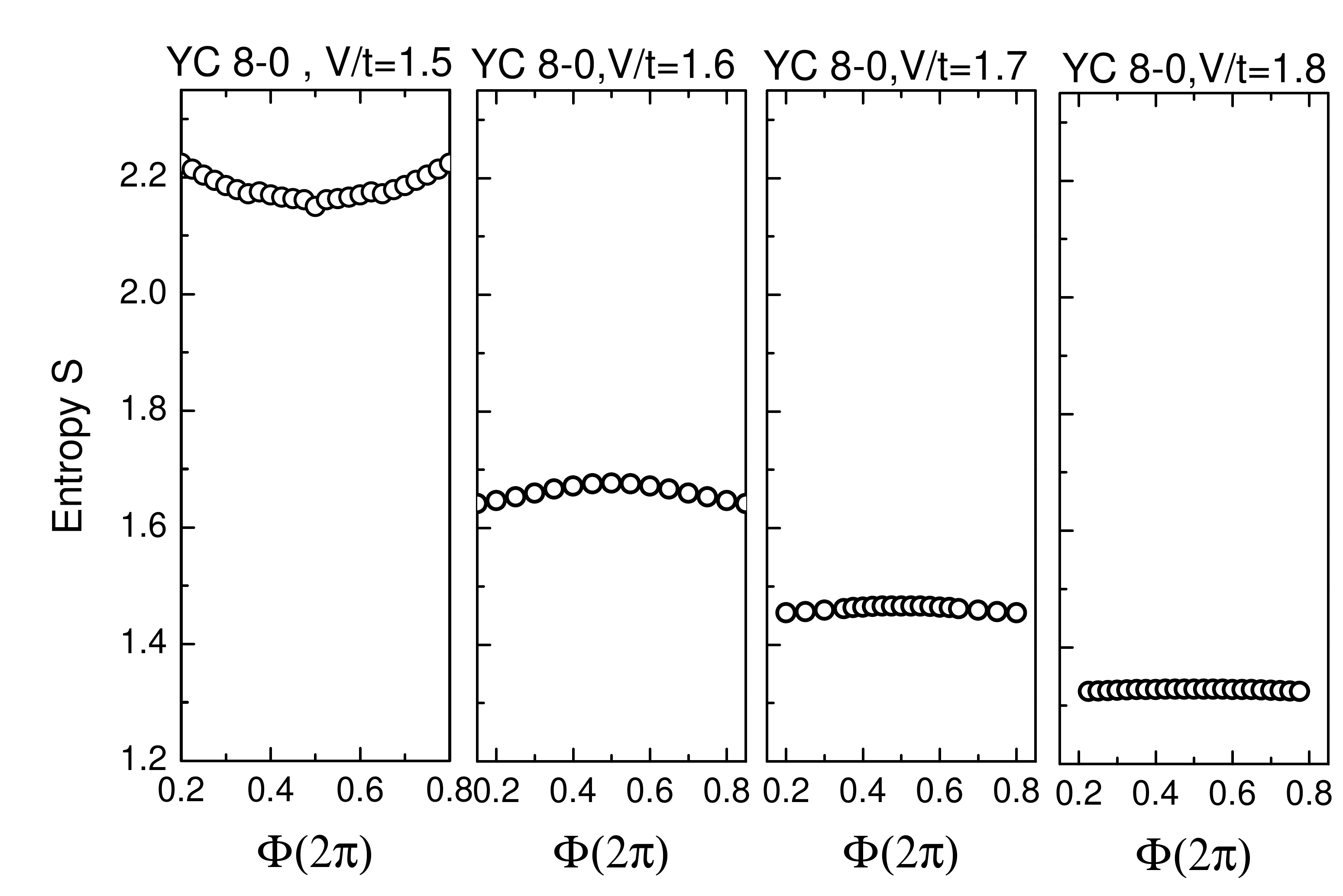}
\caption{{\bf Entanglement in the charge ordered phase}.
Entanglement entropy dependence on the external flux in the gapped charge density wave phase, $V>V_c\approx 1.3t$, for
the square lattice $\pi-$flux model introduced in the main text.}  \label{fig:gapped}
\end{figure}

\section{Entanglement entropy in the gapped phase}
In the main text, we have focused on the non-trivial scaling behavior of the
entanglement dependence on the external flux.
The strong dependence of the EE on the external flux constitutes a fingerprint of
the gapless Dirac cone structure.
In this section, we analyze the situation where the Dirac fermions acquire a gap. %become gapped after the phase transition happens.
In the $\pi$-flux and honeycomb models, this occurs when the interaction is strong enough $V\!>\! V_c$ ($V_c$ is phase transition point).

In Fig.~\ref{fig:gapped}, we show the DMRG data for the EE in the charge density wave phase ($V\!>\! V_c$).
We observe that the EE has little dependence on the flux $\Phi$,
in contrast to the gapless Dirac semimetal occuring at $V\!<\! V_c$.
This can be understood from the fact that once the system is sufficiently deep in the gapped phase,
its correlation length will be smaller than the circumference, %typical system size,
and most quantities should be hardly influenced by the twisted boundary conditions.
Thus, the EE of the insulating phase is expected to become more insensitive to the flux as the gap increases,
which is akin to Thouless's picture of localization in which the energy spectral flow of insulators is robust against boundary conditions.
Moreover,  in the charge density
wave phase $V\! >\! V_c$, the EE does not follow the scaling function any more.
These results show that the scaling behavior observed at $V\!<\! V_c$ is
tied to the gapless Dirac fermions.

%\end{widetext}

\end{appendices}

\end{widetext}

\end{document}